\documentclass[conference]{IEEEtran}
\usepackage{cite}
\usepackage{amsmath,amssymb,amsfonts}
\usepackage{listings}
\usepackage{graphicx}
\usepackage{xcolor}
\usepackage{tabularx}
\usepackage{tabulary}
\usepackage{array}
\usepackage{booktabs}
\usepackage{verbatim}
\usepackage{pifont}             
\usepackage[shortlabels]{enumitem} 	
\usepackage{makecell} 
\usepackage[bookmarks=false]{hyperref}
\usepackage{algorithmic}

\usepackage{tikz}
\usetikzlibrary{calc} 
\usetikzlibrary{arrows.meta}
\usetikzlibrary{positioning}

\usepackage[linesnumbered,ruled,commentsnumbered]{algorithm2e}

\tikzset{
	>=Latex,
	line/.style={draw,->},
	anode/.style={rectangle,draw,
		align=center,rounded corners,minimum height=4em,font=\strut},
	bnode/.style={anode,fill=white, font=\strut},
	cnode/.style={anode, fill=cyan!20, font=\strut},
treenode/.style = {circle,
	draw=black,thick, fill=white, align=center, minimum size=1cm},
root/.style     = {treenode, font=\footnotesize},
env/.style      = {treenode, font=\footnotesize}, 
dummy/.style    = {circle,draw}
	
}

\usepackage{xcolor} 
\definecolor{light-gray}{gray}{0.95}

\begin{document}

\title{Toward a Generic Mapping Language \\
for Transformations between\\ RDF and Data Interchange Formats}



\author{
\IEEEauthorblockN{
    Aljosha Köcher,
    Artan Markaj,
    Alexander Fay
}
\IEEEauthorblockA{
Institute of Automation Technology\\
Helmut Schmidt University, Hamburg, Germany\\
Email: aljosha.koecher@hsu-hh.de, artan.markaj@hsu-hh.de, alexander.fay@hsu-hh.de\\}
}

\maketitle

\begin{abstract}
While there exist approaches to integrate heterogeneous data using semantic models, such semantic models can typically not be used by existing software tools. Many software tools --- especially in engineering --- only have options to import and export data in more established data interchange formats such as XML or JSON. 
Thus, if an information which is included in a semantic model needs to be used in a such a software tool, automatic approaches for mapping semantic information into an interchange format are needed.
We aim to develop a generic mapping approach that allows users to create transformations of semantic information into a data interchange format with an arbitrary structure which can be defined by a user. 
This mapping approach is currently being elaborated. In this contribution, we report our initial steps targeted to transformations from RDF into XML. At first, a mapping language is introduced which allows to define automated mappings from ontologies to XML. Furthermore, a mapping algorithm capable of executing mappings defined in this language is presented. 
An evaluation is done with a use case in which engineering information needs to be used in a 3D modeling tool.
\end{abstract}


\section{Introduction} 
\label{sec:Introduction}
\emph{Semantic Web Technologies}, such as ontologies---which define relevant terms of a domain as well as the relations between those terms in a formal manner---were originally intended to solve the problem of vast amounts of unrelated data in the \emph{World Wide Web} \cite{BHL_TheSemanticWeb_2001}.
Since then, Semantic Web Technologies have been increasingly used to represent complex interrelations which typically occur in technical contexts, e.g., during the life cycle of \emph{Cyber-Physical Systems} (CPS) \cite{HKK+_OntologyBuildingforCyberPhysical_2020} or in industrial plant engineering \cite{ESS+_Ontologybaseddataintegrationin_2017b}.
Engineering of industrial plants takes place in a multi-disciplinary environment in which various engineering domains and disciplines collaborate. 
Typically, every discipline has its own set of tools and databases, resulting in heterogeneous information. 
Semantic Web Technologies are considered promising to link and integrate this information into a single machine-readable, semantic model \cite{ESS+_Ontologybaseddataintegrationin_2017b}.

While there are efforts to make use of semantic technologies for shared and formal models in engineering, existing software tools in industrial practice lack support for an integration of semantically modeled information. 
Typically, data integration is done using less formal interchange formats such as the \emph{Extensible Markup Language} (XML) or \emph{JavaScript Object Notation} (JSON). 
In contrast, semantically modeled information is encoded using \emph{Resource Description Framework} (RDF) triples, while RDF Schema (RDFS) and the \emph{Web Ontology Language} (OWL) are used to define a schema on top of RDF data \cite{HJS_SemanticWebArchitecture_2011}.
In order to use semantic information within software tools, such information needs to be transformed into an established interchange format. 
Furthermore, the challenge is not only that interchange formats are required, but importing tools depend upon certain data models. Thus, a configurable transformation approach is important.

With this research, we aim to create a generic mapping language to transform RDF data into data interchange formats. In this contribution, the following initial findings are presented: (a) A mapping language along with (b) a mapping algorithm that allow to map information contained in an RDF graph to XML. The mapping language allows for user-defined mappings so that the exact transformation of information into an XML structure can be controlled by a user.

The remainder of this contribution starts with a short overview on related approaches to transforming semantic models in Section \ref{sec:relatedWork}. In Section \ref{sec:concept}, both the mapping language and an algorithm to execute mappings are presented. Ultimately, this contribution concludes with a summary and outlook in Section \ref{sec:summary}.

\section{Related Work} 
\label{sec:relatedWork}
An early approach to transform OWL to XML-based \emph{Computer Aided Engineering Exchange} (CAEX) is presented in \cite{RFB_KonvertierungvonOWLPlanungsergebnissennach_2010}. In this contribution, the need to transform ontological information into an exchange format used by engineering tools is recognized. Thus, the approach of the authors in \cite{RFB_KonvertierungvonOWLPlanungsergebnissennach_2010} has the same goal as our approach. Unfortunately, their contribution does not describe a general-purpose mapping language, but a mapping algorithm that includes a static mapping logic that cannot be changed by user-defined mapping rules. Furthermore, the approach is only suitable for CAEX as a target format and not all elements of an ontology can be transferred.

Hua and Hein build on \cite{RFB_KonvertierungvonOWLPlanungsergebnissennach_2010} and present a bidirectional approach that can transform information from OWL ontologies to CAEX and vice versa \cite{HuHe_InterpretingOWLComplexClasses_2019}. The approach is very extensive and covers large parts of the semantics of OWL, even complex class expressions. However, the approach doesn't contain a general-purpose mapping language for transferring information from OWL into hierarchical exchange formats either.

\emph{Extensible Stylesheet Language Transformations} (XSLT) as defined in \cite{Mic_XSLTransformationsXSLTVersion_2017} would allow more general transformations from OWL ontologies into all kinds of XML documents. However, XSLT is a transformation language between two XML documents so it would require mappings to be created against the XML-serialized form of an ontology. This is significantly more cumbersome than working against the actual OWL model and impedes the use of powerful ontology features like querying and reasoning. Furthermore, we strive to create a mapping language that can --- at a later point --- also generate other interchange formats such as JSON. Thus, XSLT is not a suitable foundation for general-purpose mappings from an ontological model to existing data interchange formats. 

The \emph{RDF Mapping Language} (RML) is a powerful mapping language to transform data contained in interchange formats such as JSON or XML into RDF. RML was originally published in \cite{DVC+_RML:AGenericLanguage_2014b}. RML is defined as a superset of the \emph{RDB to RDF Mapping Language} (R2RML). RML has been further developed since 2014 and is now available as a W3C Draft with comprehensive documentation and tooling. Unfortunately, RML is a monodirectional mapping language that supports only mappings from existing data formats into RDF. Thus, RML is inherently not suited for transformations from RDF to data interchange formats.
Nevertheless, the approach presented in this paper follows the rationale behind RML, and we aim to create a comparable mapping language for a reversed transformation from RDF to well-established data interchange formats.

\section{Concept}
\label{sec:concept} 
\begin{figure}
    \centering
    \includegraphics[width=0.9\linewidth]{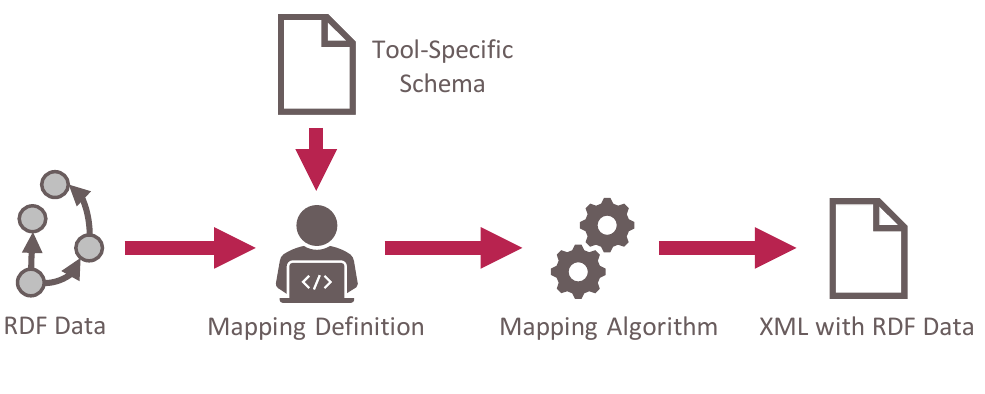}
    \caption{Overview of the concept.}
    \label{fig:conceptOverview}
\end{figure}

The main idea of this contribution is to allow users to create custom mapping rules which transform certain parts of an ontology into a data interchange format such as XML (see Figure~\ref{fig:conceptOverview}). 
Such rules allow engineering knowledge to be kept in an ontology and to transform only certain information of this ontology into an exchange format.
In this section, the mapping language and algorithm are introduced by means of our current achievement: a transformation from RDF to XML.

To illustrate the concepts described in this contribution, a robot use case is taken as a simplified running example. From different robot configurations, the one most suitable for a given task needs to be found. For simplicity, we only consider arm length as a parameter to select. The arm length depends on various factors of the robot application, e.g., the expected load as well as desired size of the working area.
We assume that all these dependencies are already maintained in an ontology.

Let's assume that manufacturing of other parts of a robot cell is done not by the robot manufacturer, but instead by other companies which need a 3D model of the exact robot configuration with all specific parameters such as the arm lengths. 
With a parametric 3D design tool, a 3D model for one specific configuration can typically be derived from a base model by importing a set of parameters.
This parameter information is contained in an ontology and needs to be automatically imported as an XML file into a 3D design tool. 

\begin{figure*}[ht]
    \centering
    \scalebox{0.60}{
        \begin{tikzpicture}
        [env/.style={circle, draw=gray!60, fill=gray!10, thick, minimum size=10mm}]
			\begin{scope}[
			every node/.style = {font=\large, thick}, 
			every label/.style = {font=\large, align=center}]
				\node[env, label={\text{ex:RobotConfiguration\_ABC}}] (A) at (-4,0) {};
				\node[env, label={\text{ex:Parameter1}}] (B) at (2, 0) {};
				\node[ ] (B1) at (7,1) {"arm1"};
				\node[ ] (B2) at (7,-0.5) {200};
				\node[env, label={\text{ex:Parameter2}}] (C) at (3,-2) {};
				\node[ ] (C1) at (8,-1.5) {"arm2"};
				\node[ ] (C2) at (8,-2.5) {260};
				
				\node[env, label={\text{ex:Parameter3}}] (D) at (-9,-1.5) {};
				\node[ ] (D1) at (-14,-1) {"arm3"};
				\node[ ] (D2) at (-14,-2.2){220};
			\end{scope}
			
			\begin{scope}[>={Stealth[black]},
				every node/.style={fill=white},
				every edge/.style={draw=black, font=\large}]
				every label/.style={font=\normalsize}
				\path [->] (A) edge node {\texttt{ex:hasParameter}} (B); 
				\path [->] (A) edge[bend right=20] node [pos=0.5] {\texttt{ex:hasParameter}} (C); 
				\path [->] (A) edge[bend left=20] node[pos=0.4] {\texttt{ex:hasParameter}} (D);
				\path [->] (B) edge[] node[pos=0.5] {\texttt{ex:hasName}} (B1);
				\path [->] (B) edge[] node[pos=0.5] {\texttt{ex:hasValue}} (B2);
				\path [->] (C) edge[] node[pos=0.5] {\texttt{ex:hasName}} (C1);
				\path [->] (C) edge[] node[pos=0.5] {\texttt{ex:hasValue}} (C2);
				\path [->] (D) edge[] node[pos=0.5] {\texttt{ex:hasName}} (D1);
				\path [->] (D) edge[] node[pos=0.5] {\texttt{ex:hasValue}} (D2);
			\end{scope}
		\end{tikzpicture}}
    \caption{Simplified excerpt from an exemplary ontology containing a robot configuration with various parameters (arm lengths in this case). Whether these elements exist in a \emph{materialized} (i.e., persistent) form or are created trough a reasoning process is not important.}
    \label{fig:robotConfigurationInOntology}
\end{figure*}
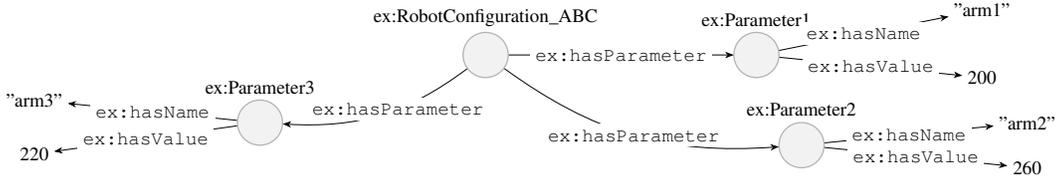

The objective is therefore to generate an XML-based parameter description of a certain robot configuration. During engineering, a robot is designed using an ontology, so that all parameters might be modeled as shown in Figure \ref{fig:robotConfigurationInOntology}. It is important to note that this information may either be stored in a persistent manner or can be inferred by reasoning and may, thus, only be available during query execution. As long as the information can be queried with a SPARQL query, there is no difference for our mapping approach. This is a benefit of working with an actual ontology model instead of its serialization --- which would be required by XSLT.

\subsection{Mapping Language}
The developed mapping language is based on RML and uses a comparable vocabulary.
A mapping file contains definitions written in the developed vocabulary using one of the RDF serialization formats, e.g., Turtle. 
The output document is an XML file containing information of an ontology in a desired structure. The output document may initially be empty or may already contain valid XML. Subsequently, the components of the mapping vocabulary will be explained using the robot example.

\begin{lstlisting}[
    basicstyle=\footnotesize,
    caption={Example mapping definition to map parameters from Figure \ref{fig:robotConfigurationInOntology} into the XML form of Listing \ref{lst:parameterXml}},
    label=lst:parameterMapping]
@prefix ol: <http://www.hsu-hh.de/aut/ontologies/olif#>.
...
<#ParameterMapping> a ol:DataMap;
  ol:ontologicalSource [
    ol:source "parameters.ttl";
    ol:sourceType ol:File;
    ol:queryLanguage ol:Sparql;
    ol:query 
      "PREFIX ex: <http://.../example#> 
      SELECT ?parameterName ?parameterValue WHERE {
        ex:RobotConfiguration_ABC a ex:RobotConfiguration;
          ex:hasParameter ?parameter.
        ?parameter ex:hasName ?parameterName;
          ex:hasValue ?parameterValue.
        }"
  ];
  ol:container "/parameters";
  ol:snippet 
   "<ParamWithValue>
      <name>length_${parameterName}</name>
      <typeCode>mm</ typeCode>
      <value >${parameterValue} mm</value>
    </ParamWithValue>".
\end{lstlisting}

Prefix declarations are used for abbreviating URIs. 
Comparable to RML's \verb|TriplesMap|, a \verb|ol:DataMap| is the class of all mapping definitions that define a mapping from RDF source elements to XML. It describes which concepts of an RDF graph should be mapped to XML. In the case of the robot arm a \verb|<#ParameterMapping>| is defined.

Every \verb|ol:DataMap| has to have one individual related via \verb|ol:ontologicalSource| which defines all source elements that are part of a mapping. This individual further specifies:
\begin{enumerate}[(a)] 
    \item The actual source of the ontology to map. 
	\item The type of the source, which may be either a file or a SPARQL endpoint
    \item The technology used to find source information within the source ontology. Currently, only SPARQL is supported.
    \item A SPARQL query whose results are used to create elements in the output file.
\end{enumerate}

For our running example, we use a parameters.ttl file as the source document. The SPARQL query of Listing~\ref{lst:parameterMapping} searches for all parameters of a specific robot configuration \verb|ex:RobotConfiguration_ABC| and returns the names and values of all parameters found.

While an \verb|ol:ontologicalSource| defines the source, i.e. the elements to be mapped, \verb|ol:container| and \verb|ol:snippet| define how these source elements will be represented in the output document.
An \verb|ol:container| is an XPath expression defining XML elements that the elements to be mapped will be inserted into. The elements targeted might already exist in an output document so that elements can be mapped into an existing XML structure. If they don't exist, they are created during mapping. An \verb|ol:container| may use results of the source's SPARQL query.
An \verb|ol:snippet| represents the actual XML structure that will be created through the current mapping rule. It is an XML string that may also use results of the SPARQL query. In our example, the variables \verb|parameterName| and \verb|parameterValue| are used to dynamically create XML representations for each parameter returned by the query.

\subsection{Mapping Algorithm}
In order to execute mapping definitions given in the language presented in the previous subsection, an initial mapping algorithm has been developed.
The steps necessary to transform an RDF graph into an XML file are shown in the pseudo code of Algorithm~\ref{alg:mappingAlgorithm} and are explained in this subsection.

\begin{algorithm}
	\label{alg:mappingAlgorithm}
	\caption{Mapping Algorithm}
	\SetKwInOut{Input}{Input}
	\SetKwInOut{Output}{Output}
	\Input{Path to mapping definition file $P_M$, \\ Path to XML file $P_X$}
	\Output{XML structure to write to $P_X$}
	\BlankLine
	$M_M \leftarrow$ Create mapping Model from $P_M$\;
	$X \leftarrow$ Open or create XML document from $P_X$\;
	$D \leftarrow$ Get all mapping definitions inside $M_m$\;
	\ForEach{$d \in D$}{
		$M_S \leftarrow$ Get model from source\;
		$Q \leftarrow$ Get query\;
		$R \leftarrow$ Execute $Q$ against $M_S$\;
		$C \leftarrow$ Insert all $r \in R$ into container XPath expression\;
		$N_C \leftarrow$ Execute XPaths to obtain or create container node for each $c \in C$\;
		\ForEach{$n \in N_C$}{
			$S \leftarrow$ Insert all $r \in R$ into snippet\; 
			\ForEach{$s \in S$}{
				\textbf{Append} $s$ to $n$\;
			}
		Update or create $n$ in $X$
		}
	}
	Write updated $X$ to $P_x$
\end{algorithm}

The input of the algorithm is a file with mapping definitions as described in the previous subsection as well as a file path to an XML file which will later contain the results of all mapping defintions.
First, a \emph{mapping model} $M_M$ is created from the mapping file with path $P_m$ using the Jena Framework\footnote{https://jena.apache.org/} so that the mapping definitions can be easily queried using SPARQL (\textbf{Line 1}).
Afterwards, the given XML file path $P_X$ is opened or an empty XML file is created in case the path doesn't exist (\textbf{Line 2}).

All mapping definitions are retrieved using a SPARQL query against the mapping model (\textbf{Line 3}). In Listing \ref{lst:parameterMapping}, only one definition is shown, but there may be an arbitrary number of definitions inside a mapping file.
The following steps are then executed for every mapping definition inside the mapping file.

If a mapping definition's source is given as a file, this file is loaded as a so-called \emph{source model} $M_S$ to also run SPARQL queries against it. In case a SPARQL endpoint is given, no further actions are required as the source model can directly be queried (\textbf{Line 5}).
The \verb|ol:query| found inside the current mapping definition is executed against the current source model and the results are stored in $R$ (\textbf{Lines 6-7}). 
Then, the \verb|ol:container| is taken and any occurring SPARQL variable inside the container expression is filled with its corresponding value of the query results $R$. 
As there can be multiple SPARQL results, this may lead to multiple XPath container definitions $C$ (\textbf{Line 8}). 
Every container XPath is then executed against the output document which may result in multiple container nodes $N_C$ being found (or created if they do not yet exist) for every single container definition.

The \verb|ol:snippet| needs to be added into each found container node (\textbf{Lines 10-16}).
To achieve that, eventually occurring variables inside the snippet (${parameterName}$ and ${parameterValue}$ in Listing \ref{lst:parameterMapping}) need to be resolved using the current results $R$. This may result in multiple instantiations of a snippet. Every instantiation is inserted into every container node (\textbf{Lines 12-14}). After that, the changed container node containing the snippets is updated or appended to the XML document (\textbf{Line 15}).
After all mapping definitions are executed, the updated or created document is written to the given path $P_X$ (\textbf{Line 18}).

\smallskip

Listing \ref{lst:parameterXml} shows the part of the XML output generated for the ontology shown in Figure \ref{fig:robotConfigurationInOntology} and the mapping definition given in Listing \ref{lst:parameterMapping}. In this case, there might be other parameters already defined, so an existing XML file was used. Furthermore, the structure of a parameter must be adhered to in order for the XML to be accepted when importing it into the 3D modeling software which was used.

\begin{lstlisting}[
    label=lst:parameterXml,
    basicstyle=\footnotesize,
    caption={XML output of the example robot use case. The two parameters (third not shown) have been created from information contained in the ontology of Figure~\ref{fig:robotConfigurationInOntology} by applying the mapping definition given in Listing~\ref{lst:parameterMapping}}, 
]
<?xml version="1.0" encoding="utf-8"?>
<ParamWithValueList>
    <parameterTypes>
        ...
    </parameterTypes>
    <parameters>
        <ParamWithValue>
            <name>length_arm1</name>
            <typeCode>mm</typeCode>
            <value>200 mm</value>
        </ParamWithValue>
        <ParamWithValue>
            <name>length_arm2</name>
            <typeCode>mm</typeCode>
            <value>260 mm</value>
        </ParamWithValue>
        <!-- Third parameter not shown -->
    </parameters>
</ParamWithValueList>
\end{lstlisting}

This exemplary evaluation example has shown that the proposed transformation works. A systematic transformation from ontologies to XML as a data interchange format has been achieved. The generated XML contains dynamic data from an ontology in a structure specified by user-defined mapping definitions.

\section{Summary \& Outlook} 
\label{sec:summary}
Current research in automated ontology transformations only provides approaches with fixed mapping rules for certain use cases---e.g., a transformation of an OWL ontology into a CAEX model \cite{HuHe_InterpretingOWLComplexClasses_2019}. 
Engineering tools typically do not allow for a direct integration of semantically modeled information. Thus there is a need for a generic transformation approach allowing users to specify mapping definitions to transfer semantic information into established data interchange formats with a specific data model.

In this contribution, initial steps to achieve such a generic mapping approach have been shown. A mapping language allowing transformations from RDF models into XML together with a corresponding mapping algorithm has been presented with a simple engineering use case. Both the mapping language and the mapping algorithm are published as open-source and --- together with additional documentation --- are available at https://github.com/hsu-aut/olif.


The approach presented in this contribution is generic with regards to the mapping rules that can be defined. But it is currently limited to only generating XML documents. 
In our future work, we are going to extend the language and algorithm to also allow for JSON and other exchange formats to be generated from an ontology. 
Another important aspect is a sound investigation of the expressivity of the mapping language. In future research, we plan to ensure that any given data structure in the target format can be created by using the mapping approach.
Additionally, while we have closely based our research on the approach taken by RML, we currently defined a distinct mapping language. We will look into ways of better aligning our works with the existing RML vocabulary.
And finally, we aim to conduct additional evaluations with larger use cases.

\bibliographystyle{IEEEtran}
\bibliography{references} 

\end{document}